\documentclass{article}
\usepackage{graphicx} 
\usepackage{amscd}
\usepackage{verbatim}
\usepackage{appendix}
\usepackage{hyperref}
\usepackage{amssymb}
\usepackage{amsthm}
\usepackage{dsfont}
\usepackage{amsmath}
\usepackage{xcolor}
\usepackage{tipa}
\usepackage{tabularx}
\usepackage{mathtools}

\DeclarePairedDelimiterX\braket[2]{\langle}{\rangle}{#1 \delimsize\vert #2}

\usepackage{graphicx} 
\def\be{\begin{equation}}
\def\ee{\end{equation}}
\def\ba{\begin{align*}}
\def\ea{\end{align*}}
\def\nb{\nabla}
\def\l{\lambda}

\def\lb{\label}

\def\m{\mu}
\def\n{\nu}
\def\s{\sigma}
\def\a{\alpha}
\def\b{\beta}
\def\Om{\Omega}

\def\g{\gamma}
\def\d{\delta}

\def\r{\rho}

\def\z{\zeta}

\def\pd{\partial}

\def\fr{\frac}

\begin{document}
\begin{center}
\vspace{24pt} { \large \bf Causal Structure of Higher Curvature Gravity } \\
\vspace{30pt}
\vspace{30pt}
\vspace{30pt}
{\bf Mohd Ali \footnote{mohd.ali@students.iiserpune.ac.in}}, {\bf Vardarajan
Suneeta\footnote{suneeta@iiserpune.ac.in}}\\
\vspace{24pt} 
{\em  The Indian Institute of Science Education and Research (IISER),\\
Pune, India - 411008.}
\end{center}
\date{\today}
\bigskip
\begin{center}
{\bf Abstract}
\end{center}
In this paper, we analyze the causal structure of Generalized Quadratic Gravity (GQG) and Einsteinian Cubic Gravity (ECG). It is well known that gravitons in higher-curvature theories can exhibit superluminal propagation, rendering the conventional definition of causal structures based on null curves inadequate. Instead, the causal structure must be defined using the fastest propagating modes, which travel along characteristic surfaces.
The superluminal propagation in higher-curvature theories has significant implications for black holes. Specifically, if the Killing horizon of a black hole is not a characteristic surface corresponding to the fastest propagating mode, the horizon can no longer function as a causal barrier. Our analysis demonstrates that GQG with a genuine fourth-order equation of motion possesses only null characteristics, implying that the horizon is a characteristic surface.
Furthermore, we perform a detailed characteristic analysis of ECG. We show that while all null surfaces are characteristic surfaces in ECG, the converse is not true - there exist non-null characteristic surfaces. In particular, we identify a non-null characteristic surface in a Type N spacetime in the algebraic classification of spacetimes. Despite the existence of multiple characteristic surfaces in ECG, we establish that the black hole horizon remains a characteristic surface.
\newpage
\section{Introduction}
General Relativity (GR) is one of the most elegant and experimentally validated theories, providing a robust framework for describing gravitational phenomena in the weak-field, low-energy regime \cite{Will:2014kxa}. However, its validity in strong-field regimes remains an open question, as we currently lack sufficient experimental or observational evidence to confirm its applicability under such extreme conditions. This uncertainty makes it imperative to explore extensions of GR, such as higher curvature theories, which naturally arise in many approaches to quantum gravity. It is also important to study various higher curvature gravity theories to check for deviations from GR in the gravitational wave signal from binary black hole mergers.
\\
In higher curvature theories, the equations of motion (EoM) typically have higher order derivatives than two, leading to a more intricate spectrum of propagating modes and the causal structure. These modes often have distinct propagation speeds depending on fields and polarization, with some modes propagating superluminally\cite{Reall:2014pwa, Tanahashi:2017dgr}. In the presence of such superluminal propagation, the usual GR notion of causal structure, defined with respect to null curves, becomes inadequate. Instead, the causal structure must be redefined in terms of the fastest propagating modes.
\begin{figure}[h]
  \centering
  \includegraphics[width=0.95\textwidth]{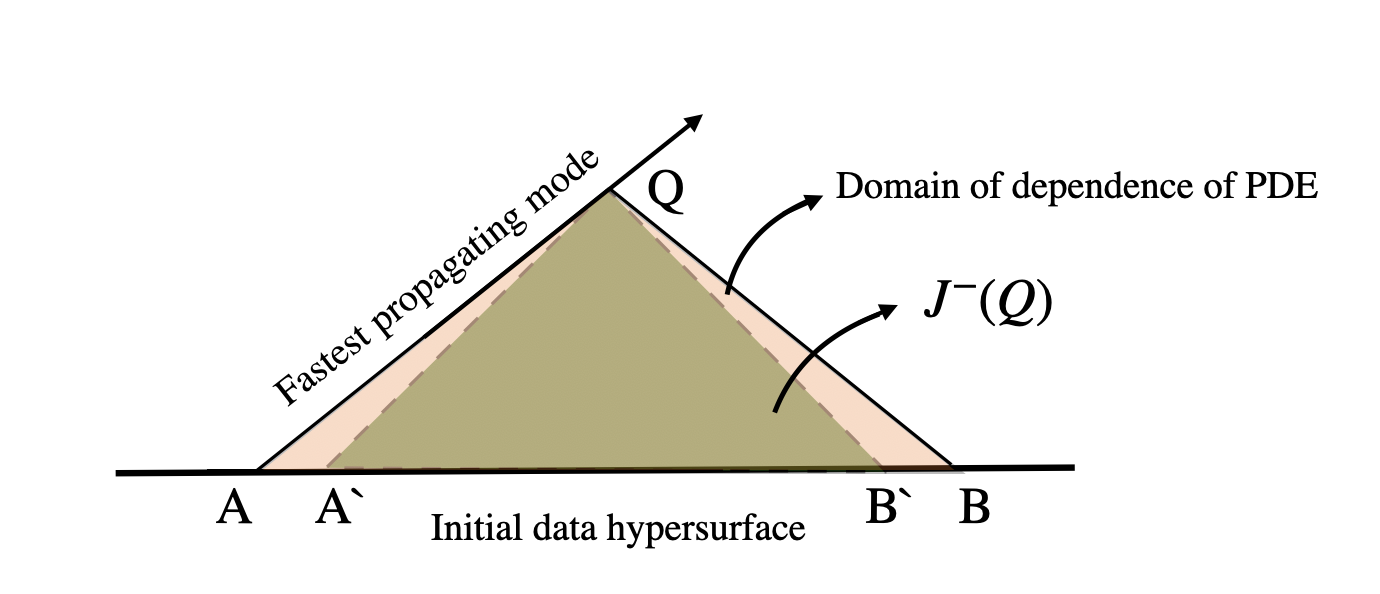}
  \caption{In this figure, the region $ABQ$ represents the domain of dependence of the interval $AB$ on the initial hypersurface, as determined by the PDE. The green region $A'QB'$ corresponds to the causal past of the spacetime point $Q$, as determined by the spacetime metric.
 }
  \label{fig:fastestmode}
\end{figure}
For many partial differential equations (PDEs), the fastest-propagating modes travel along characteristic surfaces, as shown in Figure~\ref{fig:fastestmode}. These surfaces are determined by the \emph{principal symbol} of the PDE—that is, the coefficient of the highest-order derivative terms in the EoM. A useful way to understand this is by considering the initial value problem. Suppose \( AB \) is an interval on the initial data hypersurface, and we ask: what is the maximal region in which the solution can be determined using initial data restricted to \( AB \)? If the EoM  as a PDE is of order \( n \), then the initial data must specify the field and its derivatives up to order \( n-1 \). The EoM then determines the \( n \)-th order derivatives in terms of this initial data. This evolution is possible only when the coefficient of the highest-order derivative term is invertible. The surfaces on which this coefficient vanishes mark the boundary beyond which Cauchy evolution breaks down. These are the characteristic surfaces, and because they limit the region that can be causally influenced by the initial data, the modes propagating along them are the fastest modes permitted by the PDE.
 As a result, the causal structure of higher curvature theories is encoded in the principal symbol of their governing equations.
 \\
  In GR, the fastest propagating modes travel along null curves, and therefore it makes sense to define the causal structure with respect to null curves. This is not true in generic higher curvature theory. So it is important to analyze the properties of the principal symbol in such theories. This fact has implications for black holes in higher curvature theories. Let \( M \) be the manifold representing the entire spacetime. If we extend the definition of a black hole region from GR as \( M \setminus J^-(\mathcal{I}^+) \), where \( J^-(\mathcal{I}^+) \) is the causal past of future null infinity \( \mathcal{I}^+ \), this definition inherently relies on the causal structure defined with respect to null curves associated with the spacetime metric. As we have discussed, in higher curvature theories that admit superluminal propagating modes, the causal structure is determined by the fastest propagating modes rather than the null structure of the spacetime metric. In these theories, if the horizon is not a characteristic surface corresponding to the fastest mode, the black hole boundary can no longer act as a causal barrier. This motivates us to study the causal structure of higher curvature theories.
\\
Generic higher curvature theory has lots of pathologies, like ill-posed initial value problems, instabilities, and perturbative ghosts. But there are higher curvature theories with second-order EoM and, which therefore, are free from such instabilities and ghosts, like Lovelock theories. Concerning the question of whether the event horizon is a characteristic surface, it is shown by Izumi in \cite{Izumi:2014} that in the Gauss-Bonnet theory, the Killing horizon is a characteristic surface for all polarization modes of the graviton and therefore, no modes can leak from the horizon. Reall, Tanahashi, and Way in \cite{Reall:2014pwa} have generalized this result to the full Lovelock class of theories. In Horndeski theories, gravitons see different effective metrics than photons, and the consequences for black hole thermodynamics has been discussed in \cite{sheikhjabbari}.
\\
Generally, the theories with the EoM higher than the second order have linearized ghosts and are also considered to be ill-posed. Nevertheless, there are theories with ghosts that have well-posed initial value problems and stable dynamical properties \cite{Deffayet:2023wdg, Held:2023aap, Noakes:1983xd}. This makes the higher derivative theories more interesting and worth exploring. One of the main goals of studying these theories is to come up with some physically motivated criteria to define good gravitational theories. For example, the causality criteria of Camanho, Edelstein, Maldacena, and Zhiboedov (CEMZ) require the theory not to have a Shapiro time \textit{advance} \cite{Camanho:2014apa}.
Edelstein, Ghosh, Laddha, and Sarkar in \cite{Edelstein:2021jyu, Edelstein:2024jzu} have shown that Generalized Quadratic Gravity (GQG) in a shock wave background for some class of couplings has Shapiro time delay and therefore satisfies CEMZ causality criteria.
\\
{\bf Summary of results and outline:}\\
In this paper, we have analyzed the causal structure in two higher curvature theories: one containing quadratic in curvature terms in the Lagrangian, and another that also contains cubic terms.
The first of these theories is Generalized quadratic gravity (GQG) in arbitrary dimensions, whose Lagrangian has arbitrary linear combinations of squares of the Riemann tensor, Ricci tensor and Ricci scalar. The second theory for which we have studied the causal structure is Einsteinian Cubic Gravity (ECG),
a special cubic curvature theory that has only a massless graviton in the spectrum when linearized about a maximally symmetric background.
The motivation for studying GQG is that it gives the general form of $O(\alpha'^{2})$ (where $\alpha'$ is the square of string length) corrections to Einstein gravity in string theory. For certain specific values of constants multiplying the curvature terms, the theory becomes second order (Gauss-Bonnet gravity) and the causal structure of that theory has already been analyzed. In this paper, we analyze all the other cases, when the equations of motion are fourth order. The characteristics analysis of linearized perturbations in some higher derivative theories which are fourth order has been previously studied for some specific backgrounds in \cite{Benakli:2015qlh}.

In GQG, we find that except for some special values of couplings, the characteristic surfaces are always null with respect to the spacetime metric. Thus, in particular, if there is an event horizon in the spacetime, it is a characteristic surface as it is null.

We have also analyzed ECG, which is a specific higher curvature gravity theory with cubic in curvature terms in the Lagrangian. In this theory, any null hypersurface is a characteristic surface. There are also non-null characteristic surfaces possible. However, we show that any black hole spacetime whose near-horizon metric is of the form of a product of maximally symmetric spaces has the event horizon as a characteristic surface. The derivation of these results shows that this is a special feature of ECG, and a \emph{general} cubic gravity theory will not have this property.

To study the causal structure, we have used the method of characteristics. In particular, we have looked at the principal symbol, which encodes the information about the characteristic surfaces.
In Section (\ref{I}), we consider a theory with fourth-order EoM, described by a diffeomorphism invariant Lagrangian. Following Reall in \cite{Reall:2021}, we define the principal symbol and study its symmetries, as well as the implications of the action principle and diffeomorphism invariance on the principal symbol. The symmetries of the principal symbol are used in subsequent sections to analyze charateristic surfaces. In Section (\ref{II}), as a warm-up example, we analyze the characteristics of the Riemann-squared theory in \(D\)-dimensions and the implications for Killing horizons in this theory. In Section (\ref{III}), we examine GQG in \(D\)-dimensions and study its characteristics when the theory has genuinely a fourth-order EoM. In Section (\ref{IV}), we present a characteristic analysis of Einsteinian Cubic Gravity. First, we analyze ECG on an arbitrary background but are only able to analyze the null characteristics. We also discuss black hole horizons. To analyze the non-null case, we perform a characteristics analysis on Type N spacetimes in the Weyl classification. In Section (\ref{V}), we discuss the results and outline future directions.
\section{Principal Symbol and its Symmetries}\lb{I}
Let us consider theories described by the action below with metric $g_{\m\n}$ and matter $\Phi_I$.
\be \lb{I1}
S[g_{\m\n},\Phi_I]= \fr{1}{16\pi G}\int d^Dx \sqrt{-g}L(g,\Phi_I).
\ee
The gravitational EoM obtained from the above action is
 \be \lb{I2}
E^{\m\n}_g=-\fr{-16\pi G}{\sqrt{g}}\fr{\d S}{\d g_{\m\n}}=0.
 \ee
Similarly, the matter EoM is
 \be \lb{I3}
E_{m}^I=-\fr{-16\pi G}{\sqrt{g}}\fr{\d S}{\d \Phi_I}=0
 \ee
where $g$ denotes metric and $m$ denotes matter. Let us assume that the EoM is fourth order, then the variation of EoM takes the following form,
 \be \lb{I4}
\d E^{\m\n}_g=M_{gg}^{\m\n \a\b,\g\r\d\z}\hspace{1mm} \pd_\g\pd_\r\pd_\d\pd_\z \d g_{\a\b}+ M_{gm}^{\m\n J,\g\r\d\z }\hspace{1mm} \pd_\g\pd_\r\pd_\d\pd_\z  \d \Phi_J +...
 \ee
 \be \lb{I5}
\d E^I= M^{IJ,\g\r\d\z}_{mm} \hspace{1mm} \pd_\g\pd_\r\pd_\d\pd_\z \d \Phi_J +M^{I\a\b,\g\r\d\z}_{mg} \hspace{1mm}\pd_\g\pd_\r\pd_\d\pd_\z \d g_{\a\b} +...
 \ee
 where $M_{gg}^{\m\n\a\b,\g\r\d\z}$, $M_{gm}^{\m\n J,\g\r\d\z }$, $M^{IJ,\g\r\d\z}_{mm} $ and $M^{I\a\b,\g\r\d\z}_{mg}  $ denote the coefficients of the highest derivatives of the metric and matter fields in the equations of motion (EoM), which we refer to as the principal tensors. The ellipses in the above equation denote the terms with less than four derivatives. We define the matrix of the principal symbol for the EoM by contracting the  ${\g\r\d\z }$ indices of $M$ with an arbitrary covector $K_\m$ as
 \be\lb{I6}
\pmb{\mathcal{P}(K)}= \begin{pmatrix}
M_{gg}^{\m\n \a\b,\g\r\d\z}\hspace{1mm} K_\g K_\r K_\d K_\z& M_{gm}^{\m\n J,\g\r\d\z }\hspace{1mm} K_\g K_\r K_\d K_\z\\
 M^{I\a\b,\g\r\d\z}_{mg} \hspace{1mm} \hspace{1mm}K_\g K_\r K_\d K_\z & M^{IJ,\g\r\d\z}_{mm} \hspace{1mm} \hspace{1mm}K_\g K_\r K_\d K_\z
\end{pmatrix}
 \ee

 which we can write as,
 \be \lb{I7}
\pmb{\mathcal{P}(K)}= \begin{pmatrix}
P_{gg}^{\m\n \a\b}(K)& P_{gm}^{\m\n J}(K)\\
  P^{I\a\b}_{mg}(K) & P^{IJ}_{mm}(K)
\end{pmatrix}
 \ee
 where $P_{AB}^{****}(K)=M_{AB}^{****,\g\r\d\z}\hspace{1mm} K_\g K_\r K_\d K_\z$ with $A,B\in \{g,m\}$, is the principal symbol of the field $B$ coming from $E_{A}^{**}$\footnote{* here is a proxy for the index}. The principal symbol matrix $\pmb{\mathcal{P}(K)}$ is an endomorphism on the vector space $V_{pol}$ of “polarization” vectors of
the  form $\mathbf{T}=(T_{\m\n}, T_I )$ where $T_{\m\n}$ is symmetric. As we will see, $\pmb{\mathcal{P}(K)}$ is always degenerate due to diffeomorphism invariance, but then we can define a gauge equivalence class of polarizations, which we can call "physical polarization." The covector  $K_\m$  is called a characteristic covector if there exists a non-zero $\mathbf{T}$  in the gauge equivalence class of polarizations such that it satisfies the characteristic equation $\pmb{\mathcal{P}(K)}\textbf{T}=0$. Furthermore, a hypersurface with a normal covector that is a characteristic covector everywhere on the hypersurface is called a characteristic hypersurface. On these surfaces, the coefficient of the highest derivative in (\ref{I4}) and (\ref{I5}) is non-invertible. Therefore, one cannot express the highest derivative in the EoM in terms of lower derivatives. Thus, characteristic surfaces represent the boundary of the Cauchy development. There can be multiple characteristics, and the modes propagating along the outermost characteristic surfaces are the fastest-moving modes.  Consequently, information about the causal structure of a theory is encoded in the principal symbol of the EoM of the theory.
\subsection{Symmetries of Principal symbol}\lb{IA}
In this section, we will follow \cite{Reall:2021} to find the principal symbol's symmetries. It is clear from the (\ref{I2})–(\ref{I5})that,
\be \lb{IA1}
M_{gg}^{\m\n \a\b,\g\r\d\z}=M_{gg}^{(\m\n )\a\b,\g\r\d\z}=M_{gg}^{\m\n (\a\b),\g\r\d\z}=M_{gg}^{\m\n \a\b,(\g\r\d\z)}
\ee
\be \lb{IA2}
M_{gm}^{\m\n J,\g\r\d\z }=M_{gm}^{(\m\n )J,\g\r\d\z }=M_{gm}^{\m\n J,(\g\r\d\z )}
\ee
\be \lb{IA3}
 M^{IJ,\g\r\d\z}_{mm} = M^{IJ,(\g\r\d\z)}_{mm}
\ee
\be \lb{IA4}
M^{I\a\b,\g\r\d\z}_{mg}=M^{I(\a\b),\g\r\d\z}_{mg} =M^{I\a\b,(\g\r\d\z)}_{mg}
\ee
\subsubsection{Implication of action principle on Principal symbol}\lb{IAA}
Let $g_{\m\n}(x)$ and $\Phi_I(x)$ be any background field configuration (need not be a solution to the EoM) in the configuration space. Let $g_{\m\n}(x,\l_1,\l_2)$ and $\Phi_I(x,\l_1,\l_2)$ be the two-parameter family of field configurations in the configuration space such that  $g_{\m\n}(x,0,0)=g_{\m\n}(x)$ and $\Phi_I(x,0,0)=\Phi_I(x)$, and $g_{\m\n}(x,\l_1,\l_2)$, $\Phi_I(x,\l_1,\l_2)$ agree with $g_{\m\n}(x,0,0)$, $\Phi_I(x,0,0)$ outside a compact set . We will denote derivatives with respect to $\l_1$ and $\l_2$ as $\d_1$ and $\d_2$ respectively. It can easily be shown that
\begin{multline}\lb{IAA1}
\d_2\d_1 S=-\fr{1}{16 \pi G}\int d^Dx \sqrt{-g}\Big[(E^{\m\n}_g\d_2\d_1g_{\m\n}+E^I_{m}\d_2\d_1\Phi_I)\\
+(M_{gg}^{\m\n \a\b,\g\r\d\z}\hspace{1mm} \pd_\g\pd_\r\pd_\d\pd_\z \d_2 g_{\a\b}+ M_{gm}^{\m\n J,\g\r\d\z }\hspace{1mm} \pd_\g\pd_\r\pd_\d\pd_\z  \d_2 \Phi_J +...)\d_1g_{\m\n}\\
+(M^{IJ,\g\r\d\z}_{mm} \hspace{1mm} \pd_\g\pd_\r\pd_\d\pd_\z \d_2 \Phi_J +M^{I\a\b,\g\r\d\z}_{mg} \hspace{1mm}\pd_\g\pd_\r\pd_\d\pd_\z \d_2 g_{\a\b} +...)\d_1\Phi_I\Big]
\end{multline}
where ellipses involve terms with less than four derivatives in total acting on the variation of fields. In order to get the above equation, we have used (\ref{I4}) and (\ref{I5}), and the fact that the variation is compactly supported, and therefore, we can neglect the boundary terms. Since the above equation is covariant and we are not interested in terms with fewer than four derivatives, we can replace partial derivatives with covariant derivatives in the equation above. Further, by integrating by parts twice
\begin{multline}\lb{IAA2}
\d_2\d_1 S=-\fr{1}{16 \pi G}\int d^Dx \sqrt{-g}\Big[(E^{\m\n}_g\d_2\d_1g_{\m\n}+E^I_{m}\d_2\d_1\Phi_I)\\
+(M_{gg}^{\m\n \a\b,\g\r\d\z}\hspace{1mm}\nb_\g\nb_\r\d_1g_{\m\n} \nb_\d\nb_\z \d_2 g_{\a\b}+ M_{gm}^{\m\n J,\g\r\d\z }\hspace{1mm} \nb_\g\nb_\r\d_1g_{\m\n}\nb_\d\nb_\z  \d_2 \Phi_J +...)\\
+(M^{IJ,\g\r\d\z}_{mm} \hspace{1mm}\nb_\g\nb_\r\d_1\Phi_I \nb_\d\nb_\z \d_2 \Phi_J +M^{I\a\b,\g\r\d\z}_{mg} \hspace{1mm}\nb_\g\nb_\r\d_1\Phi_I\nb_\d\nb_\z \d_2 g_{\a\b} +...)\Big]
\end{multline}
Now antisymmetrizing the variation in $\l_1$ and $\l_2$ and then computing the equation at the background configuration gives,
\begin{multline}\lb{IAA3}
0=-\fr{1}{16 \pi G}\int d^Dx \sqrt{-g}\Big[(M_{gg}^{\m\n \a\b,\g\r\d\z}-M_{gg}^{\a\b\m\n ,\g\r\d\z})\nb_\g\nb_\r\d_1g_{\m\n} \nb_\d\nb_\z \d_2 g_{\a\b}\\
+( M_{gm}^{\m\n J,\g\r\d\z }- M_{mg}^{J\m\n ,\g\r\d\z })\nb_\g\nb_\r\d_1g_{\m\n}\nb_\d\nb_\z  \d_2 \Phi_J\\
+(M^{IJ,\g\r\d\z}_{mm}-M^{JI,\g\r\d\z}_{mm})\nb_\g\nb_\r\d_1\Phi_I \nb_\d\nb_\z \d_2 \Phi_J \\
+(M^{I\a\b,\g\r\d\z}_{mg} -M^{\a\b I,\g\r\d\z}_{gm} )\nb_\g\nb_\r\d_1\Phi_I\nb_\d\nb_\z \d_2 g_{\a\b}+...\Big]
\end{multline}
The above expression has to hold for an arbitrary compactly supported variation, which implies
\begin{align}\lb{IAA4}
    M_{gg}^{\m\n \a\b,\g\r\d\z}&=M_{gg}^{\a\b\m\n ,\g\r\d\z}\hspace{1cm} M_{gm}^{\m\n J,\g\r\d\z }&= M_{mg}^{J\m\n ,\g\r\d\z } \hspace{1cm}M^{IJ,\g\r\d\z}_{mm}&=M^{JI,\g\r\d\z}_{mm}
\end{align}
which for covector $K_\m$ gives,
\begin{align}\lb{IAA4}
    P_{gg}^{\m\n \a\b}(K)&=P_{gg}^{\a\b\m\n}(K)\hspace{1cm} P_{gm}^{\m\n J }(K)&= P_{mg}^{J\m\n }(K) \hspace{1cm}P^{IJ}_{mm}(K)&=P^{JI}_{mm}(K)
\end{align}
Thus, the principal symbol is symmetric. We emphasize that one can always derive the above equation from the action principle for any action whose EoM is of even order. For the odd-order EoM, the action principle will not give a symmetric principal symbol.
\subsubsection{Implication of Diffeomorphism invariance on the principal symbol}\lb{1B}
As we know, under diffeomorphisms with compact support, the action is invariant.  In a diffeomorphism invariant theory, the diffeomorphism invariance implies the Bianchi identity,
\be\lb{IB1}
\nb_{\m}E^{\m\n}_g-L^\n_I E^I_m=0
\ee
where $L^\n_I$ is the coefficient of the infinitesimal change in $\Phi_I$ under an infinitesimal diffeomorphism  \footnote{If $\Phi_I$ is a n-tensor field then  $L_{\n I}= \fr{\delta (\mathcal{L}_\xi \Phi_I)}{\delta \xi^\n}$, where $\mathcal{L}_\xi$ is the Lie derivative with respect to vector field $\xi^\n$ .}. This must hold for an arbitrary configuration of fields. Using (\ref{I4}) and (\ref{I5}) we can write (\ref{IB1}) as
\be \lb{IB2}
M_{gg}^{\m\n \a\b,\g\r\d\z}\hspace{1mm}\pd_\m\pd_\g\pd_\r\pd_\d\pd_\z  \d g_{\a\b}+ M_{gm}^{\m\n J,\g\r\d\z }\hspace{1mm} \pd_\m\pd_\g\pd_\r\pd_\d\pd_\z \d \Phi_I+...=0
\ee
where the ellipsis denotes terms with fewer than four derivatives. Since the above equation is true for an arbitrary configuration, the coefficient of the highest derivative must vanish.
\begin{align}\lb{IB3}
M_{gg}^{\n(\m| \a\b|,\g\r\d\z)}&=0 \hspace{1cm} M_{gm}^{\n(\m| J|,\g\r\d\z)}=0
\end{align}
where $(\m| \a\b|,\g\r\d\z)$ means that upon fixing $\a\b$, it is symmetric in $\m\g\r\d\z$. For an arbitrary covector $K_\m$, the above equation implies that
\begin{align}\lb{IB4}
K_\m P_{gg}^{\m\n \a\b}(K)&=0 \hspace{1cm} K_\m P_{gm}^{\m\n J}(K)=0
\end{align}
This relation will hold for any higher curvature theory because it is just an outcome of the diffeomorphism invariance of the action.
\section{Riemann squared theory}\lb{II}
For illustration of the method, in this section, we are interested in studying the causal structure of a theory with the following action:
\be \lb{II1}
S=\int \sqrt{-g}\hspace{1mm}d^Dx (R+ \l R_{\m\n\a\b}R^{\m\n\a\b})
\ee
where $\l$ is the coupling constant associated with a higher curvature term, $D$ is the spacetime dimension, and we will assume that $D\geq4$. In generic higher curvature theory, the analysis of causal structure based on null curves (with respect to spacetime metric) does not make sense. Therefore, we must do a characteristic analysis of the differential equation obtained from the above Lagrangian. The equation of motion (EoM) for the above Lagrangian is,
\be \lb{II2}
R_{\m\n}-\fr{1}{2}g_{\m\n}R + \l\Big( 2 R_{\m}^{~\a\b\g}R_{\n\a\b\g}+2 \nb^\a \nb^\b R_{\m\a\n\b}-\fr{1}{2}g_{\m\n} R_{\g\d\a\b}R^{\g\d\a\b}\Big)=0
\ee
It is clear from the above equation that it is a fourth-order quasi-linear PDE. Therefore, we can write the above equation as,
\be\lb{II3}
M_{\m\n}\hspace{1mm}^{\a\b,\g\r\d\z}(g)\hspace{1mm} \pd_\g\pd_\r\pd_\d\pd_\z g_{\a\b}+ O(\pd^3g)=0
\ee
where $M_{\m\n}\hspace{1mm}^{\a\b,\g\r\d\z}$ is the coefficient of the highest derivative term in the EoM. It is clear from (\ref{II2}) that the highest derivative term will come from $\nb^\a \nb^\b R_{\m\a\n\b}$, and it will only depend on the metric. Let $\Sigma$ be a co-dimension 1 surface with the normal $K^\m$. We can define the principal symbol for the above PDE as
\be \lb{II4}
P_{\m\n}\hspace{1mm}^{\a\b}(x,K)=M_{\m\n}\hspace{1mm}^{\a\b,\g\r\d\z}(g)\hspace{1mm} K_\g K_\r K_\d K_\z
\ee
 We know from the method of characteristics that the fastest propagating modes are tangent to the characteristic surface. The characteristic equation for (\ref{II3}) is given by
 \be \lb{II5}
det(P(x,k))=0
 \ee
 The above equation tells us that the kernel of the principal symbol will give us the modes moving along the characteristic surface. Let $T_{\a\b}$ be a symmetric tensor corresponding to a polarization mode of the graviton. The possible $T_{\a\b}$ satisfying
 \be \lb{II6}
\l Q_{\m\n}(K,T)=P_{\m\n}\hspace{1mm}^{\a\b}(x,K)T_{\a\b}=0
\ee
will give the kernel of the principal symbol, and its dimension will give the number of modes propagating along characteristics. It can easily be shown that the $Q_{\m\n}(K,T)$ for Riemann Squared theory is,
\be\lb{II7}
 Q_{\m\n}(K,T)= -2 K^2 K^2 T_{\m\n}+2 K^2 K_{\m} K^\a T_{\a\n} +2 K^2 K_{\n} K^\a T_{\a\m} -2 K_{\m}K_{\n} K^\a K^\b T_{\a\b}
\ee
It is straightforward to check that $K^\m Q_{\m\n}(K,T)=0$ and for $T'_{\m\n}=T_{\m\n}+K_{(\m}X_{\n)}$ where $X_{\n}$ is any covector field, $Q_{\m\n}(K,T')=Q_{\m\n}(K,T)$. Therefore $K_{(\m}X_{\n)}$ is pure gauge. Now, we will split the characteristic analysis into two parts, as in \cite{Reall:2014pwa}.
\\
\\
\textbf{Null Case : $K^2=0$}
\\
\\
In this case, the equation (\ref{II6}) reduces to
\be \lb{II8}
 K_{\m}K_{\n} K^\a K^\b T_{\a\b}=0
\ee
which implies $K^\a K^\b T_{\a\b}=0$. Therefore, the characteristic equation only fixes one component of the $T_{\a\b}$, but $D$ of the components of $T_{\a\b}$ are just  pure gauge, and we can fix $D-1$ by suitable gauge choice. To see this more explicitly, we can choose a null basis $\{k^\m, l^\m, m^\m_i\}$ such that,
\be\lb{II9}
k.k=l.l=k.m_i=l.m_i=0 \hspace{1mm}\&\hspace{2mm} k.l=-1 ,\hspace{1mm}m_i.m_j=\d_{ij}
\ee
We will denote contraction with respect to $\{k^\m, l^\m, m^\m_i\}$ as $\{0,1,i\}$. Let us choose one of the null basis vectors $k=K$. The equation in \ref{II8} implies that $T_{00}=0$ ; further, in this basis, $T_{1\m}$ are pure gauge modes due to diffeomorphism invariance \footnote{In the null basis $T_{\m\n}= T_{00}l_\m l_\n + T_{11}K_{\m}K_{\n}+2T_{1i}K_{(\m}m^i_{\n)}+2T_{0i}l_{(\m}m^i_{\n)}+2T_{01}K_{(\m}l_{\n)}+ T_{ij}m^i_{(\m} m^j_{\n)}$, notice that $T_{1\m}=\{T_{11},T_{10},T_{1i}\}$ are the coefficient of the term of type $K_{(\m}X_{\n)}$, where $X_{\n}=\{K_\n,l_\n,m^i_\n\}$. Therefore they are pure gauge modes.} .
As a result, the total degeneracy of the principal symbol for the null characteristics is $\fr{D(D+1)}{2}-D-1=\fr{(D-2)(D+1)}{2}$, where $\fr{(D-2)(D+1)}{2}$ are the degrees of freedom (DoF) associated with physical propagating modes in the space of symmetric two tensors (total minus pure gauge). Notice that the degeneracy is equal to the number of DoF for massive gravitons. But it is important to remember that we are not in a $2$ derivative theory, and therefore derivatives of the metric may not be  canonically conjugate to the metric; some of them will be independent degrees of freedom. Further, we are in the eikonal limit, where all the information about the spectrum is encoded in the allowed polarization \footnote{In the eikonal limit, although the mass term is irrelevant, the polarization retains information distinguishing between massive and massless fields.}. So we cannot associate $\fr{(D-2)(D+1)}{2}$  DoF to massive gravitons directly; one must reduce the theory to second-order theory, and then these $\fr{(D-2)(D+1)}{2}$ propagating components will split into different particles that can occur in the spectrum of the theory. The crucial thing to notice is that this is true for any null surface in this theory without assuming any condition on curvature components. So, unlike the Gauss-Bonnet theory \cite{Izumi:2014} , all null surfaces are characteristic. This also tells us the Killing horizon in this theory is a characteristic surface.
\\
\\
\textbf{Non Null Case: $K^2 \neq 0$}
\\
\\
In the non-null case, it is always possible to write $T_{\m\n}=\hat{T}_{\m\n} + K_{(\m}X_{\n)}$,  for some $X$ and where $\hat{T}_{\m\n}$ is transverse, i.e $K^\m \hat{T}_{\m\n}-\frac{1}{2}\hat{T} K_b=0 $\footnote{ $\hat{T}$ is the trace of $\hat{T}_{\a\b}$}. Using the fact that $K_{(\m}X_{\n)}$ is pure gauge and $\hat{T}_{\a\b}$ is transverse, we can simplify the expression in (\ref{II6}) for the non-null case:
\be \lb{II10}
Q_{\m\n}(K,T)= -2 K^2 K^2 \hat{T}_{\m\n}+ K^2 K_{\m} K_\n \hat{T}=0.
\ee
As we want $Q_{\m\n}(K,T)=0$, its trace will also vanish, implies
\be \lb{II11}
Q(K,T)=-K^2 K^2 \hat{T}=0.
\ee
Since $K^2 \neq 0$, this implies $\hat{T}=0$, and putting it back to the equation (\ref{II10}), we get
\be \lb{II12}
Q_{\m\n}(K,T)= -2 K^2 K^2 \hat{T}_{\m\n}=0.
\ee
The only solution to the above equation is $\hat{T}_{\m\n}=0$. When $K^2 \neq 0$, the symbol is not degenerate; therefore, Riemann squared theory has no non-null characteristics.
The above analysis suggests that in the Riemann-squared theory, all the characteristics are null.  Furthermore, there are $\fr{(D+1)(D-2)}{2}$ allowed polarization modes that move along the characteristic surface.
\section{Generalized Quadratic Gravity (GQG}
\begin{figure}[h]
  \centering
  \includegraphics[width=0.65\textwidth]{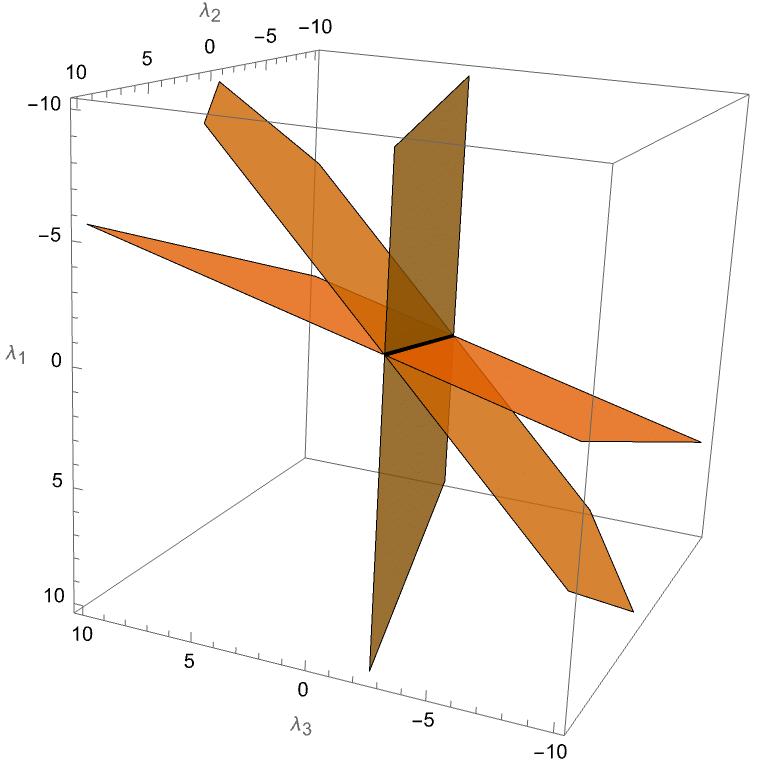}
  \caption{This figure depicts the space of couplings, with the colored regions indicating forbidden couplings.  Here, $\{\l_1, \l_2, \l_3\}$ represent the couplings in the action of GQG, with the dimension $D$ set to 5. From the above analysis, the equation for the forbidden couplings is $(\fr{1}{2}\l_2+2\l_3)(2\l_1 + \l_2 + 2\l_3)(\l_3-\l_1 +5(\l_1+\fr{1}{4}\l_2))=0$.}
  \label{fig:Coupling}
\end{figure}

\lb{III}
In this section, we are interested in studying the causal structure of theory with the following action:
\be \lb{III1}
S=\fr{1}{16\pi G}\int \sqrt{g}\hspace{1mm}d^Dx (R+ \l_3 R_{\m\n\a\b}R^{\m\n\a\b}+\l_2 R_{\m\n}R^{\m\n}+\l_1 R^2)
\ee
where $\l_i$ are the coupling constants and $D\geq 4$. The EoM for the above Lagrangian is,
\begin{multline} \lb{III2}
E_{\m\n}=R_{\m\n}-\fr{1}{2}g_{\m\n}R +  \textcolor{red}{(\l_2 +4\l_3)\square R_{\m\n}+\fr{1}{2} (\l_2+4\l_1)g_{\m\n}\square R} \\
 \textcolor{red}{-(2\l_1+\l_2+2\l_3)\nb_\m \nb_\n R} +2\l_3 R_{\a\b\g\m}R^{\a\b\g}_\n+2(\l_2+2\l_3)R_{\a\m\g\n}R^{\a\g}\\
-4\l_3R_{\m\a}R_{\n}^\a+2\l_1RR_{\m\n}-\fr{1}{2}g_{\m\n}(\l_3 R_{\m\n\a\b}R^{\m\n\a\b}+\l_2 R_{\m\n}R^{\m\n}+\l_1 R^2)=0
\end{multline}
It is evident from the above equation that the EoM is a fourth-order quasi-linear PDE of the form (\ref{II3}). The terms that contribute to the fourth-order derivative are shown in red color in the equation (\ref{III2}). Since we are interested in causal structure determined by the principal symbol, these are the only terms of relevance. For any covector $K_\m$ and symmetric 2-tensor $T_{\a\b}$, it is straight forward to show that
\begin{multline} \lb{III3}
Q_{\m\n}(K,T)=\l_1(-2K_\m K_\n K^\a K^\b T_{\a\b}+2 g_{\m\n} K^2 K^\a K^\b T_{\a\b}+2K_\m K_\n K^2 T-2 g_{\m\n}K^4T)\\
+\l_2 (-\fr{1}{2}K^4 T_{\a\b}+\fr{1}{2}K_\n K^2 T_{\m \b}K^\b+\fr{1}{2}K_\m K^2 T_{\n \b}K^\b -K_\m K_\n K^\a K^\b T_{\a\b}+\fr{1}{2}g_{\m\n}K^2 K^\a K^\b T_{\a\b}\\+\fr{1}{2} K^2 K_\m K_\n T -\fr{1}{2}g_{\m\n}K^4 T)
+\l_3(-2 K^2 K^2 T_{\m\n}+2 K^2 K_{\m} K^\a T_{\a\n} +2 K^2 K_{\n} K^\a T_{\a\m} -2 K_{\m}K_{\n} K^\a K^\b T_{\a\b})
\end{multline}
where $K^4=K^2 K^2$. We can write the above equation in terms of  $\Pi_{\m\n}(K)=g_{\m\n}K^2-K_\m K_\n$, this is $K^2$ times a projector that projects the vector onto transverse space to $K$.
 \begin{multline} \lb{III4}
Q_{\m\n}(K,T)=-2\l_1(\Pi_{\m\n}(K) \Pi^{\a\b}(K))T_{\a\b}\\
-\fr{1}{2}\l_2 (\Pi_\m^{~\a}(K) \Pi_\n^{~\b}(K)+\Pi_{\m\n}(K) \Pi^{\a\b}(K))T_{\a\b}
-2\l_3(\Pi_\m^{~\a}(K) \Pi_\n^{~\b}(K))T_{\a\b}.
\end{multline}
Another way of writing this equation is
\begin{multline} \lb{III5}
Q_{\m\n}(K,T)=\Big(-(2\l_1+\fr{1}{2}\l_2)\Pi_{\m\n}(K) \Pi^{\a\b}(K)
-(\fr{1}{2}\l_2+2\l_3)\Pi_\m^{~\a}(K) \Pi_\n^{~\b}(K)\Big)T_{\a\b}.
\end{multline}
Notice that the term inside the bracket is the principal symbol. It is evident from the above equation that that $K^\m Q_{\m\n}(K,T)=0$ and for $T'_{\m\n}=T_{\m\n}+K_{(\m}X_{\n)}$ where $X_{\n}$ is any covector, $Q_{\m\n}(K,T')=Q_{\m\n}(K,T)$. Therefore $T_{\m\n}=K_{(\m}X_{\n)}$ is pure gauge. Further, the principal symbol vanishes for any $K_\m$, when $\l_3=-\fr{1}{4}\l_2=\l_1$. This choice of couplings corresponds to the Gauss-Bonnet term in the action whose EoM is second order. Here, we restrict ourselves to theories where the equation in (\ref{III5}) is genuinely fourth order. It requires either $\l_3+\fr{1}{4}\l_2\neq0$ or $\fr{1}{4}\l_2+\l_1\neq 0$. We will assume that $\l_3+\fr{1}{4}\l_2\neq0$, as this condition will also appear in the case of the characteristics analysis for non-null case. Let us analyze the characteristic equations for GQG,
\\
\\
\textbf{Null Case: $K^2=0$}
\\
\\
In this case, the equation (\ref{III5}) reduces to,
\\
\be \lb{III6}
Q_{\m\n}(K,T)=-(2\l_1 + \l_2 + 2\l_3)K_\m K_\n K^\a K^\b T_{\a\b}=0
\ee
As we already mentioned, we don't want $2\l_1 + \l_2 + 2\l_3=0$. Otherwise, the principal symbol will be completely degenerate for $K^2=0$. Using the null basis defined in Section (\ref{II}) and the equation (\ref{III5}), we get $T_{00}=0$. Using the fact that in the null basis, $T_{1\m}$ are pure gauge modes, the total degeneracy of the principal symbol for null characteristics is $\fr{(D-2)(D+1)}{2}$. Following the logic presented in the QG, these are the number of polarizations allowed in this theory. Similar to the Riemann squared theory, this is true without putting any conditions on the curvature components.
\\
\\
\textbf{Non-null Case: $K^2 \neq0$}
\\
\\
In non-null cases, we will follow the same steps as in the last section. Without loss of generality, we can take $T_{\m\n}=\hat{T}_{\m\n}$ as the symmetric tensor that is transverse. The equation for characteristics will become $Q_{\m\n}(K,\hat{T})=0$, where we can use $K^\m\hat{T}_{\m\n}=\fr{1}{2}K_\n \hat{T}$. It can easily be seen that
\be\lb{III7}
Q(K,\hat{T})=(\l_3-\l_1 +D(\l_1+\fr{1}{4}\l_2))K^4\hat{T}=0.
\ee
Assuming that $\l_3-\l_1 +D(\l_1+\fr{1}{4}\l_2)\neq 0$, this implies $\hat{T}=0$. Putting this back into $Q_{\m\n}(K,\hat{T})=0$, we will get
\be\lb{III8}
Q_{\m\n}(K,\hat{T})=-\fr{1}{2}(\l_2+4\l_3)K^4 \hat{T}_{\m\n}=0
\ee
Since $\l_2+4\l_3\neq 0$, the above equation implies that $\hat{T}_{\m\n}=0$. In this, all the degeneracy of the principal symbol is now lifted, and therefore, there are no non-null characteristics for GQG in the considered coupling space. The allowed couplings in $D$ spacetime dimensions are
\begin{align}
    (\fr{1}{2}\l_2+2\l_3)\neq 0
    \\
    (2\l_1 + \l_2 + 2\l_3)\neq 0
    \\
    (\l_3-\l_1 +D(\l_1+\fr{1}{4}\l_2))\neq0
\end{align}
In the figure (\ref{fig:Coupling}) we have shown the forbidden couplings in dimension $D=5$.
\section{Einsteinian cubic gravity (ECG)}\lb{IV}
Recently, there has been a lot of interest in Einsteinian cubic gravity proposed by Bueno and Cano in the paper \cite{Bueno:2016}. This theory is a special higher curvature theory of gravity (cubic in curvature) defined in a way that, when linearized about a maximally symmetric background, it gives an Einsteinian spectrum. It exists for spacetime dimension $D\geq4$, and there is some evidence that it possesses Schwarzschild-like black hole solutions \cite{Hennigar:2016}. ECG does not have an Einsteinian spectrum when linearized about generic backgrounds\cite{DeFelice:2023,Bueno:2023}. The action of ECG is
\be\lb{IV1}
S = \fr{1}{16\pi G}\int d^Dx \sqrt{-g}(R-\Lambda_0+ \alpha \mathcal{L}_4 + \b \mathcal{L}_6 +\l \mathcal{P})
\ee
where
\begin{align}\lb{IV2}
\mathcal{P}=12 R_{\m \hspace{1mm}\n}^{\hspace{1mm}\r\hspace{1mm}\s}R_{\r \hspace{1mm}\s}^{\hspace{1mm}\g\hspace{1mm}\d}R_{\g\hspace{1mm}\d}^{\hspace{1mm}\m\hspace{1mm}\n} +R_{\m\n}^{\r\s}R_{\r\s}^{\g\d}R_{\g\d}^{\m\n}-12 R_{\m\n\r\s}R^{\s\r}R^{\n\s}+8R_\m^\n R_\n^\r R_\r^\m
\\
\mathcal{L}_4= \fr{1}{4}\delta^{\m_1 \m_2 \m_3 \m_4 }_{\n_1 \n_2 \n_3 \n_4 } R_{\m_1 \m_3}^{\n_1 \n_3} R_{\m_2 \m_4}^{\n_2 \n_4}
\\
\mathcal{L}_6=\fr{1}{8}\delta^{\m_1 \m_2 \m_3 \m_4 \m_5 \m_6}_{\n_1 \n_2 \n_3 \n_4 \n_5 \n_6} R_{\m_1 \m_4}^{\n_1 \n_4} R_{\m_2 \m_5}^{\n_2 \n_5} R_{\m_3 \m_6}^{\n_3 \n_6}
\end{align}
$\Lambda_0$ is the cosmological constant and $\delta^{\m_1 \m_2 ... \m_i}_{\n_1 \n_2 ...\n_i } $ is the generalized antisymmetric Kronecker delta. Notice that $\mathcal{L}_4$ and $\mathcal{L}_6$ are quadratic and cubic Lovelock terms. As we have summarized, ECG has a lot of nice properties. It is worthwhile to study the causality structure of ECG. Let us start by writing the EoM for ECG.
\be\lb{IV3}
\mathcal{E}_\m^{~\a\b\g}R_{\n\a\b\g}-\fr{1}{2}g_{\m\n}\mathcal{L}+2 \nb^\a \nb^\b \mathcal{E}_{\m\a\n\b}=0
\ee
where $\mathcal{L}=(R-\Lambda_0+ \alpha \mathcal{L}_4 + \b \mathcal{L}_6 +\l \mathcal{P})$ and $\mathcal{E}^{\m\a\n\b}=\fr{\pd \mathcal{L}}{\pd R_{\m\a\n\b}}$. We can write
\be\lb{IV4}
\mathcal{E}^{\m\a\n\b}= \a \mathcal{E}^{\m\a\n\b}_4 + \b \mathcal{E}^{\m\a\n\b}_6 + \l \mathcal{E}^{\m\a\n\b}_\mathcal{P}
\ee
where $\mathcal{E}^{\m\a\n\b}_4$ and $\mathcal{E}^{\m\a\n\b}_6$ are terms coming from Lovelock terms in Lagrangian and $\mathcal{E}^{\m\a\n\b}_\mathcal{P}$  comes from $\mathcal{P}$. Notice that the highest derivative term in (\ref{IV3}) will come  from $\nb^\a \nb^\b \mathcal{E}_{\m\a\n\b}$. Further, it is well-known that
\be\lb{IV5}
\nb_\a \nb_\b \mathcal{E}^{\m\a\n\b}_4=\nb_\a \nb_\b \mathcal{E}^{\m\a\n\b}_6=0
\ee
since $\mathcal{E}_4$ and $\mathcal{E}_6$ are coming from Lovelock terms. Now the highest derivative contribution will come from $\nb_\a \nb_\b \mathcal{E}^{\m\a\n\b}_{\mathcal{P}}$ and it will be fourth order.  It can easily be shown that
\begin{multline} \lb{IV6}
\l \mathcal{E}_{\a\b\m\n}^{\mathcal{P}}= 6\l \Big(R_{\a\n}R_{\b\m}-R_{\a\m}R_{\b\n}+g_{\b\n}R_\a^{\hspace{1mm}\s} R_{\m\s}-g_{\a\n}R_\b^{\hspace{1mm}\s} R_{\m\s}-g_{\b\m}R_\a^{\hspace{1mm}\s} R_{\n\s}+g_{\a\m}R_\b^{\hspace{1mm}\s }R_{\n\s}\\
-g_{\b\n}R^{\s\g} R_{\a\s\m\g}+g_{\b\m}R^{\s\g} R_{\a\s\n\g}+g_{\a\n}R^{\s\g} R_{\b\s\m\g}\\-3 R_{\a\hspace{1mm}\n}^{\hspace{2mm}\s\hspace{1mm}\g}R_{\b\s\m\g}-g_{\a\m}R^{\s\g}R_{\b\s\n\g} +3 R_{\a\hspace{1mm}\n}^{\hspace{2mm}\s\hspace{1mm}\g}R_{\b\s\n\g} +\fr{1}{2}R_{\a\b}^{\hspace{3mm}\s\g}R_{\m\n\s\g}\Big).
\end{multline}
Using the above equation and with a bit of algebra, we can compute the principal symbol acting on the symmetric two-tensor $T_{\m\n}$ for the EoM in (\ref{IV3}),\\

\begin{multline}\lb{IV7}
Q_{\a\b}(T,K)=P_{\a\b}^{\hspace{3mm}\m\n}(x,K)T_{\m\n}=12 K^2 K^\m K^\n R_{\m\n}T_{\a\b}+6K^2 K^\m K^\n R_{\b \m}T_{\a\n} -6 K^4 R_\b^\s T_{\a\s}\\ -18 K_\b K^\s K^\m K^\n R_{\s\n}T_{\a\m}+6 K_\b K^2 K^\m R_\m^\n T_{\a \n}+ 6K^2 K^\m K^\n R_{\b\m\n\s}T_\a^\s +6 K^2 K^\m K^\n R_{\a\m}T_{\b\n}\\ \hspace{3mm} -6K^4 R_\a ^\s T_{\b\s}-18 K_\a K^\s K^\m K^\n R_{\s\m}T_{\b\n}+6K_\a K^2 K^\m R_\m^\n T_{\b\n} +6 K^2 K^\m K^\n R_{\a\m\n\s}T^\s_\b\\ -18 K^\s K^\m K^\n K^\r R_{\a\n\b\r}T_{\s\m}-18K_\b K^\s K^\m K^\n R_{\a\m\n\r}T_\s^\r -18K_\a K^\s K^\m K^\n R_{\b\m\n\r}T_\s^\r \\-6 K^2 K^\m K^\n R_{\a\b} T_{\m\n} +6K_\b K^2 K^\m R_\a^\n T_{\m\n} +6K_\a K^2 K^\m R_\b^\n T_{\m\n}-6 K_\a K_\b K^2 R^{\m\n}T_{\m\n}\\+12 K^2 K^\m K^\n R_{\a\n\b\s}T_{\m}^\s +12 K^2 K^\m K^\n R_{\a\s\b\n}T_{\m}^\s +18 g_{\a\b}K^\s K^\m K^\n K^\r R_{\s\m}T_{\n\r}\\-12 g_{\a\b}K^2 K^\m K^\n R_\m^\r T_{\n\r}+6 g_{\a\b} K^4 R^{\m\n}T_{\m\n}+ 6 K^4 R_{\a\b}T-6K_\b K^2 K^\m R_{\a\m}T\\-6K_\a K^2 K^\m R_{\b\m}T+18 K_\a K_\b K^\m K^\n R_{\m\n} T-12 K^4 R_{\a\n\b\m}T^{\m\n}+12 K_\b K^2 K^\n R_{\a\m\n\r}T^{\m\r}\\ +12 K_\a K^2 K^\m R_{\b\n\m\r}T^{\n\r}-18 K_\a K_\b K^\m K^\n R_{\m\s\n\r}T^{\s\r}-12 g_{\a\b}K^2 K^\m K^\n R_{\m\n} T\\+ 6 K^2 K^\m K^\n R_{\a\m\b\n}T+6 g_{\a\b}K^2 K^\m K^\n R_{\m\r\n\s}T^{\r\s}
\end{multline}
where $K_\m$ is an arbitrary covector . It can easily be checked that the above principal symbol satisfies the Bianchi identity $K^\a Q_{\a\b}(T,K)=0$ and is invariant under pure gauge transformations. Furthermore, it  can be checked that for maximally symmetric spacetimes, the principal symbol vanishes as expected. This is related to the fact that when ECG is linearized about a maximally symmetric background, its EoM is almost that of the Einstein gravity\cite{Bueno:2016}.
\\
As it must be clear from the expression of the principal symbol, the ECG will not have an Einsteinian spectrum about a generic background. We also want to emphasize that unlike GQG, the principal symbol here depends on the curvature tensors, and therefore, in principle, it can lead to very different causal structures depending on the background. Hence, doing characteristic analysis is extremely important. Let us start with null characteristics.
\\
\\
\textbf{Null Case:$ K^2=0$}
\\
\\
Again, we will use the same null basis that we have defined in the section (\ref{II}). By using this null basis and putting $K^2=0$, the principal symbol simplifies.
\begin{multline}\lb{IV8}
    Q_{\a\b}(T,K)=-18 T_{\a0} R_{00}K_\b-18 T_{\b0} R_{00}K_\a -18 T_{00}R_{\a0\b0}-18 T_0^\r R_{\a00\r}K_\b-18K_\a T_0^\r R_{\b00\r}\\+18g_{\a\b}T_{00}R_{00}+18K_\a K_\b TR_{00}-18 K_\a K_\b R_{0\m0\n}T^{\m\n}=0
\end{multline}
Let us write the above equation in component form,
\begin{align}\lb{IV9}
     Q_{00}(T,K)&=0, \hspace{3mm}  Q_{01}(T,K) =0, \hspace{3mm}  Q_{0i}(T,K)=0
\end{align}
\be \lb{IV10}
 Q_{11}(T,K)= 18 T_{ii}R_{00}-18R_{0i0j}T_{ij} =0
\ee
\be \lb{IV11}
 Q_{1i}(T,K)=-18T_{i0}R_{00}-18T_{00}R_{10i0}-18T_0^\r R_{i00\r} =0
\ee
\be \lb{IV12}
 Q_{ij}(T,K)=18\d_{ij}T_{00}R_{00}-18 T_{00}R_{i0j0} =0
\ee
The equation (\ref{IV9}) is due to the Bianchi identity. Further notice that $T_{1\m}$ does not appear anywhere, which is consistent with the fact that it is pure gauge. Assuming $T_{00}=T_{0i}=0$, the equations (\ref{IV11}) and (\ref{IV12}) are trivially satisfied.  Therefore, we are able to satisfy the null characteristic equation for any $T_{\m\n}$ satisfying $T_{00}=T_{0i}=0$ and the equation (\ref{IV10}). This implies that the degeneracy of the principal symbol is $\fr{D(D-3)}{2}$. This is the same as the DoF of a massless graviton in $D$ dimensional spacetime. Therefore, any null surface is a characteristic surface in ECG. We want to emphasize that our analysis is for any spacetime for which the fourth derivative terms do not vanish for all $T_{\m\n}$. Now, let us analyze the case of a Killing horizon. As we know, at the Killing horizon, $R_{i0j0}=R_{0ijk}=0$. Using this fact, we can easily show that all the equations for the null characteristics are satisfied for all $T_{ab}$ in the equivalence class of symmetric tensors up to gauge. This implies that the principal symbol is degenerate for all \( T_{ab} \), and therefore, the dynamics is governed by a lower-order differential equation. We emphasize that this holds true for any Killing horizon. For any such solution which is also asymptotic to a maximally symmetric spacetime, the order of the differential equation interpolates from 2 in the asymptotic region, to 4 in the bulk, and less than 4 on the horizon.
\\
Furthermore, it is well known in the case of ECG that for spacetimes of the form \( M_{D'} \times M_{D-D'} \), where \( D' < D \), and \( M_{D'} \) and \( M_{D-D'} \) are maximally symmetric spaces, the equations of motion for gravitational perturbations reduce to a linearized Einstein equation with an effective Newton's constant \( G_{\text{eff}} \) \cite{Canoth}. This indicates that, in all such spacetimes, null surfaces are characteristic surfaces.
\\
This observation has significant implications for black holes in this theory. It is well known that there exists a large class of black holes where the near-horizon geometry has this product form. Consequently, for all such black holes, the horizon is a characteristic surface. This means that none of the propagating modes can classically escape or leak through spacelike paths from inside the horizon.
\\
\\
\textbf{Non-null Case: $K^2 \neq 0$}
\\
\\
In this case, we can use the fact that $T_{\a\b}$ can be split into pure gauge and transverse parts, as in Section (\ref{II}). Using the transverse property of $T_{\a\b}$ we can show
\begin{multline}\lb{IV13}
    Q_{\a\b}(T,K)=12 K^2 R_{\m\n}K^\m K^\n T_{\a\b} -6 K^4 R_\b ^\m T_{\a\m}+6 R_\n ^\m T_{\a\m}K^\n K_\b K^2+ 6K^2 R_{\b\m\n\r}T_\a^\r K^\m K^\n \\ \hspace{2.2cm}-6 K^4 R_\a^\m  T_{\m\b}+6K^2 R_{\a\m\n\r}T_\b^\r K^\m K^\n- 6 K_\a K_\b K^2 R^{\m\n}T_{\m\n}+6g_{\a\b}K^4 R^{\m\n}T_{\m\n}\\  \hspace{2.2cm} 3K^4 R_{\a\b}T-12 K^4 R_{\a\m\b\n}T^{\m\n}+12 K_\b K^2 R_{\a\m\n\r}T^{\m\r}K^\n +12K_\a K^2 R_{\b\m\n\r}T^{\m\r}K^\n\\  \hspace{2.2cm}-18 K_\a K_\b R_{\m\r\n\s}T^{\r\s}K^\m K^\n-9g_{\a\b}K^2 R_{\m\n}K^\m K^\n T +9 K^2 R_{\a\m\b\n}K^\m K^\n T \\+6 g_{\a\b}K^2 R_{\m\r\n\s}T^{\r\s}K^\m K^\n  + 6K^2 K_\a K^\n R_{\n\s}T^\s_\b
\end{multline}
Unlike GQG, the principal symbol of ECG depends on the Riemann curvature tensor, making it challenging to analyze non-null cases on arbitrary backgrounds. Therefore, we focus on analyzing it in the Ricci-flat Type N spacetime in the algebraic classification of spacetimes using the Weyl tensor.
\subsection{ECG in Type N Spacetimes}\lb{IVA}
As is well known from the algebraic classification of the spacetimes, the type N spacetime is the simplest spacetime with a nontrivial Riemann tensor \cite{Pravda:2005, Durkee:2010}. Let us introduce a null basis $\{l^\m,n^\m,m_i^\m\}$ such that,
\be\lb{IVA1}
n.n=l.l=n.m_i=l.m_i=0 \hspace{1mm}\&\hspace{2mm} n.l=1 ,\hspace{1mm}m_i.m_j=\d_{ij}
\ee
The spacetime is type N iff in some null basis; the Riemann tensor takes the following form,
\be\lb{IVA2}
R_{\a\b\m\n}=4 \Om_{ij}  l_{[\a}m_{\b]}^i l_{[\m}m_{\n]}^j
\ee
where $\Om_{ij}$ is a $(D-2)$x$(D-2)$ matrix. If $\Om$ is traceless, then the spacetime will be Ricci flat, which is the case of interest to us. Now, we will write the equation (\ref{IV7}) in a Ricci flat type N spacetime. Further, using the Ricci flatness and the fact that the contraction of Weyl curvature with itself vanishes in type N spacetime, it can easily be shown that it is a solution to ECG. It is more convenient to work in the basis $\{K^\a,l^\a,m^\a_i\}$. It can be shown that the principal symbol in this case is
\begin{multline}\lb{IVA3}
    Q_{\a\b}(T,K)=P_{\a\b}^{\hspace{3mm}\m\n}(x,K)T_{\m\n}
    \\\hspace*{2cm}=2 \Om_{ij}\Big\{-3 K^2 (K.l)^2m_\b^i T_\a ^j -3 K^2 (K.l)^2m_\a^i T_\b ^j-9(K.l)^2 m_\a^i m_\b^j T_{00}\\ \hspace{2.5cm}
   + 12 K^2 (K.l) m_\b^j m_\a^i T_{10}- 6 m_\a^i m_\b^j K^4 T_{11}-6K^2 (K.l)l_\b m_\a^j T_0^j +9 K_\b m_\a^i(K.l)^2 T_0^j\\
   \hspace{2.5cm}+6K^4 m_\a^i l_\b T_1^j
    -6K_\b m_\a^i K^2 (K.l)T_1^j-6K^2 T_0^j m_\b^i (K.l)+9 K_\a m_\b^i (K.l)^2 T_0^j\\
   \hspace{2.5cm} +6K^4 l_\a m_\b ^i T_1^j-6K^2 (K.l)K_\a m_\b^i T_1^j-6K^4 l_\a l_\b T^{ij}+6K^2 (K.l)K_\b l_\a T^{ij}\\
 \hspace{2.3cm}  +6K_\a l_\b K^2 (K.l)T^{ij}-9 K_\a K_\b (K.l)^2 T^{ij}+3g_{\a\b}K^2 (K.l)^2 T^{ij}+3K^2 (K.l)^2 m_\a^i m_\b^j T\Big\}
\end{multline}
where $\{0,1,i\}$ indexes in the above equation are contractions with respect to $\{K^\a,l^\a,m^\a_i\}$ respectively. We can always write $K_\a$ as a linear superposition of $n_\a$ and $l_\a$. It can easily be checked that the above equation satisfies the Bianchi identity and is invariant under pure gauge transformation, i.e. $T_{\a \b}\rightarrow T_{\a\b}+ K_{(a}X_{b)}$, for arbitrary $X_\a$. As shown earlier in this section, we can analyze the null case $K^2=0$ of ECG in an arbitrary background. In this part, we will focus on the non-null case, using a Ricci flat type N spacetime.
\\
Let $K^2\neq 0$, which means that $K_a$ is a non-trivial superposition of $l_\a$ and $n_\a$ and therefore $K.l \neq0$. Since $K^2\neq 0$, we can decompose $T_{\a\b}$ into a transverse part and pure gauge. The invariance of the equation (\ref{IVA3}) under pure gauge transformations allows us to choose $T_{\a\b}$ transverse, i.e $K^\a T_{\a\b}=\fr{1}{2}K_\b T$. Further, for solving the equation $Q_{\a\b}(T,K)=0$ for $T_{\a\b}$, we will assume $T_{11}=T_1^{~i}=0$. With this assumption and the transverse property of $T_{\a\b}$, the characteristics equation reduces to
\begin{multline}\lb{IVA4}
    Q_{\a\b}(T,K)=P_{\a\b}^{\hspace{3mm}\m\n}(x,K)T_{\m\n}
    \\\hspace*{2cm}=2 \Om_{ij}\Big\{-3 K^2 (K.l)^2m_\b^i T_\a ^j -3 K^2 (K.l)^2m_\a^i T_\b ^j+ 6 K^2 (K.l)^2 m_\b^j m_\a^i T
   \\
   \hspace{2cm}-6K^4 l_\a l_\b T^{ij}+6K^2 (K.l)K_\b l_\a T^{ij} +6K_\a l_\b K^2 (K.l)T^{ij}-\frac{9}{2}K^2(K.l)^2 m_\a^i m_\b^j T\\
 \hspace{2cm}  -9 K_\a K_\b (K.l)^2 T^{ij}+3g_{\a\b}K^2 (K.l)^2 T^{ij}+3K^2 (K.l)^2 m_\a^i m_\b^j T\Big\}=0
\end{multline}
Now, we will solve the characteristics equation component by component. It can easily be checked that $Q_{0\b}(K,T)=0$ and
\be\lb{IVA5}
Q_{1\b}(K,T)=18 \Om_{ij} T^{ij} (K.l)^2 \Big(K^2 l_\b- K_\b K.l\Big)
\ee
Since we want to solve for $Q_{1\b}(K,T)=0$, this implies either $\Big(K^2 l_\b- K_\b K.l\Big)=0$ or $\Om_{ij} T^{ij}=0$. If $\Big(K^2 l_\b- K_\b K.l\Big)=0$, then contraction with respect to $l^\b$ both sides and using $l.l=0$ implies $K.l=0$, which contradicts the fact that $K_\a$ is not parallel to $l_\a$. Hence $Q_{1\b}(K,T)=0$ implies $\Om_{ij} T^{ij}=0$. If we consider $T_{ij}$ as a $(D-2)$ x $(D-2)$ matrix, then this condition is the same as tr$(\Om T)=0$, where tr is a trace in the transverse directions. Now, we are left to solve for the characteristics equations in $(D-2)$ transverse directions. Let $p$ and $q$ be the directions in the null basis which are orthogonal to $l^\a$ and $n^\a$. Then,
\be \lb{IVA6}
Q_{pq}(K,T)=-6K^2 (K.l)^2\Big(T_p^{~j} \Om_{jq}+T_q^{~j} \Om_{jp}-\frac{3}{2}T\Om_{pq} \Big)
\ee
where $T$ is the trace $T_\a^\a$. Since $T_{\a\b}$ is transverse, it can easily be shown that $T_\a^\a= 2T_{01} (K.l)^{-1}$ and $T_{ii}=0$\footnote{Using the fact that $K$ is a linear combination of $n$ and $l$, we can write $K^\a=n^\a K.l + l^\a K.n$. Since, $T_{11}=0$, it can easily be shown that $n^\a T_{\a1}=T_{01} (K.l)^{-1}$. Further, $T= 2 n^\a T_{\a1} +T_{ii}= 2T_{01} (K.l)^{-1} +T_{ii}$. Now, using the transverse condition, $T_{01}=T K.l$, implying $T_{ii}=0$.}.  In order to obtain the above equation, we used the fact that  $\Om_{ij} T^{ij}=0$. The characteristics equation $Q_{pq}(K,T)=0$ implies
\be\lb{IVA7}
T_{pj} \Om_{jq}+T_{qj} \Om_{jp}=\frac{3}{2}T_{01}(K.l)^{-1}\Om_{pq}
\ee
Since all the indices take values in the transverse directions, we can write the equation (\ref{IVA7}) in $(D-2)$ x $(D-2)$ matrix notation as
\be\lb{IVA8}
T\Om +\Om T=\frac{3}{2}T_{01}(K.l)^{-1}\Om .
\ee
$\Om$ is symmetric and therefore invertible. Multiplying both sides by $\Om$ inverse, we get,
\be\lb{IVA9}
\Om^{-1}T\Om +T=\frac{3}{2} T_{01}(K.l)^{-1}\mathds{1}_{D-2}
\ee
Taking trace on both sides and using the fact that $T_{ii}=0$, we get $T_{01}=0$. This reduces the equation (\ref{IVA8}) to
\be\lb{IVA10}
T\Om +\Om T=0
\ee
The above certainly has a nontrivial solution space. For example, in D = 4, one can use a similarity transformation to set $\Om= C_1\sigma_3$, where $C_1$ is some constant and $\sigma_3$ is the third Pauli matrix. Then it is clear that $T=C_2 \sigma_1$, where $\sigma_1$ is the first Pauli matrix, is the solution to the equation (\ref{IVA10}). Further, one can use this solution to construct solutions in higher dimensions. Therefore, we have shown that ECG has non-null characteristics and it can have superluminal propagation.
\section{Discussion}\lb{V}
In this paper, we have analyzed the causal structure of Generalized Quadratic Gravity (GQG) and Einsteinian Cubic Gravity (ECG). Firstly, we analyze the Riemann-squared theory and show that the theory only possesses null characteristics, independent of the background metric. This background-independent analysis is possible because the principal symbol in this theory does not depend on curvature. We further extend this result to genuinely fourth-order GQG. By "genuinely fourth order," we mean that the theory has a non-trivial fourth-order principal symbol in both the null and non-null cases. This condition excludes the Gauss-Bonnet theory. We demonstrate that such GQG has \(\frac{(D+1)(D-2)}{2}\) polarization modes. These modes correspond to massive spin-2, massless spin-2, and scalar fields in the spectrum.
Since we are analyzing the principal symbol of a fourth-order partial differential equation (PDE), this is equivalent to analyzing the theory in the eikonal limit, in which case, the mass term becomes irrelevant. The massive and massless polarizations get mixed, and we can only count the number of helicities allowed in the theory. Further, our results imply for black holes with Killing horizons, the Killing horizon in GQG is a characteristic surface with no polarization modes of the graviton traversing it along a spacelike path.
\\
We have also shown that for ECG, all null surfaces are characteristic surfaces. This result is demonstrated on an arbitrary background. It does not appear possible to analyze the non-null characteristics on an arbitrary background. To address the non-null case, we consider Type N spacetimes in the Weyl classification. We show that in ECG, there exist non-null characteristic surfaces. Further using the fact that, $R_{00}=R_{i0j0}=0$ on the Killing horizon, we show that the null characteristic equations are satisfied for any $T_{\m\n}$ in the equivalence class $T_{\m\n}\sim T_{\m\n}+X_{(\m}K_{\n)}$, where $X_\m$ is an arbitrary covector. This tells us that on the horizon the EoM for dynamical degrees of freedom is lower order. For ECG, it is known that the linearized EoM for gravitational perturbations has the Einstein gravity form on spacetimes which are a product \( M_{D'} \times M_{D-D'} \), where \( D' < D \), \( M_{D'} \) and \( M_{D-D'} \) are maximally symmetric spaces. For black holes whose near-horizon metric is of this product form (which is the case for static vacuum black holes in many higher curvature theories), the horizon is a characteristic surface for black holes, and therefore, no modes can escape the black hole in ECG.
\\
There are several interesting directions for further exploration. One is to investigate the well-posedness of these theories. From the perspective of classical theory, well-posedness is a fundamental criterion for a physically viable classical theory. Linearizing these theories around nontrivial backgrounds typically introduces a linearized ghost in the spectrum. If the theory admits a well-posed initial value formulation, it would be particularly interesting to examine the role of these linearized ghosts and their implications for the full non-linear theory and its quantization. Furthermore, Deser and Tekin in \cite{Deser:2002jk}, have proven the positive mass theorem for full non-linear quadratic gravity (theory with $R^2$ and $R_{\m\n}R^{\m\n}$ in action). It would be worthwhile to investigate whether a similar theorem holds in the contexts of Generalized Quadratic Gravity (GQG) and Einsteinian Cubic Gravity (ECG). Such a study could shed light on the interplay between higher-curvature corrections and gravitational stability.
\subsection{Holographic Implications}
One way to think about higher curvature theory is as the low-energy limit of a UV-complete theory of gravity. String theory is one such UV-complete quantum theory of gravity. In string theory, the semiclassical limit involves taking \( G_N \rightarrow 0 \) and the string length \( \alpha'=l_s^2 \rightarrow 0 \). In the case of AdS/CFT, this limit corresponds to \( N \rightarrow \infty \) and the 't Hooft coupling \( \lambda \rightarrow \infty \). The effective theory in the bulk is Einstein gravity plus matter, meaning all bulk fields see the same metric (the metric that describes the causal structure of the theory).

We can also define a \textit{stringy regime} where \( G_N \rightarrow 0 \) and \( \alpha' \) remains finite. In this regime, there are no quantum gravitational fluctuations, but spacetime is probed by strings rather than point particles. Perturbatively in \( \alpha' \), the stringy regime corresponds to introducing higher derivative corrections, which implies that, in principle, different bulk fields may see different metrics. On the CFT side, this corresponds to \( N \rightarrow \infty \) with \( \lambda \) finite.
\\
Now, let us consider stringy black holes (i.e., black holes in the stringy regime) that are classically stable. For such solutions, one can ask if all fields see the same horizon, namely the event horizon of the black hole. If the fields are the different polarization modes of the graviton, our analysis indicates that GQG and ECG have this property. As shown by Liu and Gesteau in \cite{Gesteau:2024rpt}, the information about the causal structure of a region in the  bulk is encoded in the associated time band algebra of operators in the boundary CFT. Gesteau and Liu propose a diagnostic of the presence of a horizon in the bulk, entirely using the boundary algebra. As already stated, this could in principle, lead to different horizons for different bulk fields. It would be interesting to investigate this diagnostic in the boundary algebras and see what it predicts for stringy horizons in the bulk at least for the different polarization modes of the graviton. In the case where we consider the  $\alpha'^{2}$ corrections to string theory, by our results and those of Izumi and Reall, Tanahashi and Way, all polarization modes see the same stringy horizon. However, our analysis of ECG shows that if we go to higher orders in $\alpha'$, this property will not be true in general --- this property is true in ECG which involves very specific values of the couplings in the cubic terms in the Lagrangian.

\section{Acknowledgements} We would like to thank Sunil Mukhi and Sudipta Sarkar for useful discussions. MA acknowledges the Council of Scientific and Industrial Research (CSIR), Government of India for financial assistance.

\end{document}